\documentstyle[prl,aps,multicol]{revtex}

\begin{document}
\title{Density matrix algorithm for the calculation of dynamical
properties of low dimensional systems}
\author{}
\author{Karen A. Hallberg}
\address{Max-Planck-Institut f\"{u}r Physik komplexer Systeme,
Bayreuther Str. 40,
Haus 16, 01187 Dresden, Germany.}
\date{\today}
\maketitle

\begin{abstract}
	I extend the scope of the density matrix renormalization
group technique developed by White to the calculation of dynamical
correlation functions. As an application and  performance evaluation
I calculate the spin dynamics of the 1D Heisenberg chain.

\end{abstract}
\pacs{}


\begin{multicols}{2}

	The density matrix renormalization group method (DMRG) as
developed recently by S. White \cite{white} is a powerful algorithm
for calculating ground state energies and static properties of
low dimensional systems.
This technique leads to highly accurate results for much larger systems than
those which can be solved by straightforward exact diagonalization.
The method has been applied successfully to several problems
such as the Haldane gap of spin-1 chains \cite{spin1}, the one-dimensional
Kondo insulator \cite{kondo} and the two-chain Hubbard model \cite{twochain}.

An effective way of extending the basic ideas of this method
to the calculation of dynamical quantities was lacking, mainly
due to the fact that it is performed in real space and it is not
possible to fix the momentum as a quantum number \cite{sorensen}.
It also involves
a strong truncation of the Hilbert space and therefore much information
of the excited states is lost.

In this paper I present a way to calculate the dynamical properties using
the DMRG method. As an application I calculate the spin dynamics of the
1D isotropic Heisenberg model. The dynamics of this model has been
studied extensively \cite{1haas,3haas,7haas,12haas,13haas,15haas,mueller}
and therefore presents a good background for comparison.

The DMRG allows for a systematic truncation of the Hilbert space
by keeping the most probable states in describing a wave function
({\it e.g.~}the ground state)
of a larger system, instead of the lowest energy states usualy kept in
previous real space renormalization techniques.
The method is very well described in Ref. \cite{white} but I shall
summarize it so as to unify notations.
A general iteration of the method consists of:
i)The effective Hamiltonian is defined for the superblock 1+2+1'+2'
(a block is a collection of sites), where the blocks 1 and 1'
come from previous iterations and blocks 2 and 2' are new added ones. It is
diagonalized to obtain the ground state $|\psi_0\rangle$
(other states could be
also kept: they are called target states).
ii) The density matrix $\rho_{ii'}=\sum_j \psi_{0,ij} \psi_{0,i'j}$ is
constructed,
where $\psi_{0,ij}=\langle i\otimes j|\psi_0\rangle $,
the states $|i\rangle $ ($|j\rangle$)
belonging to the
Hilbert space of blocks 1 and 2 (1' and 2'). The eigenstates of
$\rho$ with the  highest
eigenvalues (equivalent to the most probable states of blocks
1+2 in the ground state or in the chosen target state
of the superblock) are kept up to a certain cutoff, keeping a total of
$m$ states per block.
iii) These  states form a new reduced basis to which all the operators
have to be changed and the block 1+2 is renamed as block 1.
iv) A new block 2 is added (one site in our case) and the new superblock
(1+2+1'+2') is formed as the direct product of the states of all the blocks
(the blocks 1' and 2' are identical to blocks 1 and 2 respectively).
When more than one target state is used, {\it i.e} more than one state
is wished to be well described, the density matrix is defined as:
\begin{equation}
\rho_{ii'}=\sum_l p_l \sum_j \phi_{l,ij} \phi_{l,i'j}
\end{equation}
where $p_l$ defines the probability of finding the system in the
target
state $|\phi_l\rangle $ (not necesseraly eigenstates of the Hamiltonian).

	I want to calculate the following dynamical correlation function
at $T=0$:
\begin{equation}
C_A(t-t')=\langle\psi_0|A^{\dagger}(t) A(t')|\psi_0 \rangle ,
\end{equation}
where $A^{\dagger}$ is the Hermitean conjugate of the operator $A$, $A(t)$ is
the Heisenberg representation of $A$, and $|\psi_0 \rangle $ is the ground
state of the system. Its Fourier transform is:
\begin{equation}
C_A(\omega )=\sum_n |\langle \psi_n | A |\psi_0 \rangle |^2 \;
\delta (\omega - (E_n-E_0)),
\end{equation}
where the summation is taken over all the eigenstates $|\psi_n \rangle$ of
the Hamiltonian $H$ with energy $E_n$ and $E_0$ is the ground state energy.

Defining the Green's function
\begin{equation}
G_A(z)=\langle \psi_0 | A^{\dagger}(z-H)^{-1} A |\psi_0 \rangle,
\end{equation}
the correlation function $C_A(\omega)$ can be obtained as
\begin{equation}
C_A(\omega)=-\frac{1}{\pi}\lim_{\eta\to 0^+}{\rm Im} \; G_A(\omega+i\eta +E_0).
\end{equation}

The function $G_A$ can be written in the form of a continued fraction:
\begin{equation}
\label{eq:frac}
G_A(z)=\frac{\langle \psi_0 | A^{\dagger} A|\psi_0\rangle}
{z-a_0-\frac{b_1^2}{z-a_1-\frac{b_2^2}{z-...}}}
\end{equation}
The coefficients $a_n$ and $b_n$ can be obtained using the following
recursion equations \cite{carlos,proj}:
\begin{equation}
|f_{n+1}\rangle =H|f_n\rangle -a_n|f_n\rangle -b_n^2|f_{n-1}\rangle
\end{equation}
where
\begin{eqnarray}
|f_0\rangle &=& A|\psi_0\rangle \nonumber \\
a_n&=&\langle f_n|H|f_n\rangle/\langle f_n|f_n\rangle, \nonumber \\
b_n&=&\langle f_n|f_n\rangle/\langle f_{n-1}|f_{n-1}\rangle; \;\; b_0=0
\end{eqnarray}

	An alternative way for calculating the spectra is by means of
the Liouvillian representation of the recursion method presented above
\cite{vis,mueller}. This method leads to quasi size-independent coefficients.
In the example given below, it has been seen that the
results are the same as with the
Hamiltonian representation using Eqs.~(7) and (8) \cite{horsch}.

For finite systems the Green's function $G_A(z)$ has a finite number of
poles so only a certain number of coefficients $a_n$ and $b_n$ have to be
calculated. The DMRG technique presents a good framework to calculate such
quantities. With it, the ground state, Hamiltonian and the operator $A$
required for the evaluation of $C_A(\omega)$ are obtained. An important
feature is that the reduced Hilbert space should also describe with great
precision the relevant excited states $|\psi_n \rangle $. This is achieved by
choosing the appropriate target states.
For most systems
it is enough to consider as target states the gound state $|\psi_0\rangle$ and
the first few $|f_n\rangle $ with $n=0,1...$ and $|f_0\rangle=
A|\psi_0\rangle$ as described above. In doing so,  states
in the reduced Hilbert space
relevant to the excitated states connected to the ground state via the
operator of interest $A$ are included. The fact that  $|f_0\rangle$ is
an excellent trial state, in particular, for the lowest triplet excitations
of the two-dimensional antiferromagnet was shown in Ref.~\cite{linden}.

Another straightforward election of target states is to take the excited
eigenstates  and the ground state. But this is possible only when the
quantum number of the above mentioned states can be fixed.
 Otherwise, when
diagonalizing the Hamiltonian, the lowest lying excitations of the whole
space are obtained and not those of the appropriate symmetry sector.
With the DMRG technique we can fix the total $S^z_T$ and parity but not,
 for example, the momentum $q$ of the system.
 In the example given below, I attempted to obtain the first
excited states for a given $q$ by using $|f_0\rangle =S^z_q |\psi_0\rangle $
(where $S^z_q=\sum_j e^{iqR_j}S^z_j$)
as a trial state for the diagonalization procedure in step (i) above.
Due to the
fact that there is a small but non-zero overlap between $|f_0\rangle$ and
$|\psi_0\rangle$, the algorithm does not remain in the symmetry sector
with a given $q$.
Because of this, I found it convenient to use $|\psi_0\rangle$ and
$|f_n\rangle $ with $n=0,1..$ as target states.

Of course, if the number $m$ of states kept per block is fixed, the more
target states  considered, the less precisely each one of them are
described. An optimal number of target states
and $m$ has to be found for each case. Due to this reduction, the
algorithm can be applied up to certain lenghts, depending on the states
involved. For longer chains, the higher energy excitations will become
inaccurate. Proper sum rules have to be calculated to determine the errors
in each case.

	As an application of the method I calculate
\begin{equation}
\label{eq:szzn}
S^{zz}(q,\omega)=\sum_n |\langle \psi_n | S^z_q |\psi_0 \rangle |^2 \;
\delta (\omega - (E_n-E_0)),
\end{equation}
for the 1D isotropic Heisenberg model.

As I already mentioned,
the spin dynamics of this model has been extensively studied. The lowest
excited states in the thermodynammic limit
are the famous des~Cloiseaux-Pearson (dCP) triplets
\cite{15haas}, having total spin $S^T=1$. The dispersion of this
spin-wave branch is:
\begin{equation}
\label{eq:dcp}
\omega^l_q=\frac{J\pi}{2}|\sin (q)|
\end{equation}
Above this lower boundary there exists a two-parameter continuum of
excited triplet states that have been calculated using the Bethe ansatz
approach \cite{yam} with an upper boundary given by
\begin{equation}
\omega^u_q=J\pi|\sin (q/2)|
\end{equation}
It has been shown \cite{1haas}, however,
that there are excitations above this upper
boundary due to higher order scattering processes,
with a weight that is at least one order of magnitude lower
than the spin-wave continuum.
Based on selection rules, Bethe ansatz calculations and numerical
diagonalization of small clusters, M\"uller {\it et al.~} \cite{1haas}
proposed the following approximate expression for the out-of-plane dynamical
structure factor:
\begin{equation}
\label{eq:szz}
S^{zz}(q,\omega)=\frac{A}{\sqrt{\omega^2-{\omega^l_q}^2}}\Theta(
\omega-\omega^l_q)\Theta(\omega^u_q-\omega)
\end{equation}
where $A$ is a constant and $\Theta(x)$ a cutoff step function that
was considered so that the sum-rules are satisfied. A similar expression
for an exactly solvable
model (the Haldane-Shastry model) has been obtained \cite{haldane}.
The low energy properties of
this model and those of the nearest neighbour Heisenberg model we are
considering belong to the same universality class.

In the following I will present the numerical results.
The values for $N=20$ sites (without reduction of the Hilbert space) and
exact calculations using the Lanczos technique and exploiting all the
symmetries coincide exactly.
For larger systems I used $m=200$ states per block and periodic boundary
conditions.

In Fig. 1 I show the spectrum for various systems lenghts and $q=\pi$
and $q=\pi/2$.
The delta peaks of Eq.~(\ref{eq:szzn}) are broadened by a Lorentzian for
visualizing purposes. For this case it was enough to take 3 target states,
{\it i. e.~} $|\psi_0\rangle$, $|f_0\rangle = S^z_{\pi}|\psi_0\rangle$ and
$|f_1\rangle$.
I also plot the analytical expression (\ref{eq:szz}). There is good agreement
up to $N\simeq 40$ with the envelope of our data.
For larger values the peaks at
 $\omega /J\simeq 2$ acquire high weight, which grows with $N$. The
second peak seems to be somewhat shifted, also having a higher weight.
Due to the truncation of the Hilbert space the
spectrum of $H$ is also reduced.
I notice that if we consider only the first ($\sim 10$) coefficients
$a_n$ and $b_n$, the spectrum at low energies remains essentially unchanged.
Minor differences arise at $\omega /J\simeq 2$. This is another indication
that only the first $|f_n\rangle$ are relevant for the low energy
dynamical properties for finite systems.

In the inset of
Fig. 1 the spectrum for $q=\pi/2$ and $N=28$ is shown. For this case
I considered 5 target states {\it i. e.~} $|\psi_0\rangle$,
$|f_0\rangle = S^z_{\pi/2}|\psi_0\rangle$, $|f_n\rangle\; n=1,3$ and
$m=200$. Here, and for all the cases considered, I have verified that
the results are very weakly dependent on the weights $p_l$ of the target
states, as long as the appropriate target states are chosen.
For lenghts where this value of $q$ is not defined I took the nearest
value.
I found that with these parameters the spectrum starts developing spurious
peaks for larger systems. The coefficients also present a larger dispersion
with $N$ than for the $q=\pi$ case. To be able to go further, one should
consider larger $m$ values.

The  $q=\pi/2$ case is the most unfavourable one because it involves high
energy excitations. Although I have included some of these states as
target states, the reduced Hilbert space related to  $|\psi_0\rangle$ and
$S^z_{\pi/2}|\psi_0\rangle$ have very small overlap and many states are
needed to describe correctly both target states. For  $q=\pi$, instead,
the overlap of Hilbert spaces is very high and the target states and
low energy excitations are better described. The difference between the
$q=\pi/2$ and $q=\pi$ cases can be seen by comparing the ground state
energy, it being more precise in the latter case by a factor of
3  in the relative error for $m=200$ and $N=28$ (the relative error
of the ground state for the $q=\pi$ case is $10^{-6}$).

	Even though I am including states with a given momentum
as target states, due to the particular real-space construction of the
reduced Hilbert space, this translational symmetry is not fulfilled
and the momentum is not fixed. To check how the reduction on the Hilbert
space influences the momentum $q$ of the target state
$|f_0\rangle =S^z_q|\psi_0\rangle$, I
calculated the expectation values
\begin{equation}
\label{eq:sq}
\langle \psi_0 |S_{-q'}^z S_q^z|\psi_0\rangle
\end{equation}
for all $q'$. If the momenta of the states were well defined, this value is
proportional to $\delta_{q-q'}$ if $q\neq 0$. For $q=0$, $\sum_r S^z_r=0$.
In Fig. 2a) I show the expectation values (\ref{eq:sq}) for $q=\pi/2$
(using $S^z_{\pi/2}|\psi_0\rangle$ as one of the target states)
and different lenghts.
I see that, as the system becomes larger (higher
reduction of the Hilbert space), the $q$ value becomes less defined,
presenting a wider distribution. The figure shows a marked oscillation
in the expectation value. This is due to
the fact that the system is built from two identical blocks and that
$\left |\langle \psi_0 |S_{\pi}^z |\psi_0\rangle\right |$ is small but
non-zero ($\simeq 10^{-3}$ for the largest system).
These should disappear when using
the finite-size method \cite{white}.
The momentum distribution for $q=\pi$ is shown in Fig. 2b) in a
semilogarithmic scale. In this figure I have shifted the values by
.003 so as to have well-defined logarithms. I have also neglected the
points where Eq.~(\ref{eq:sq}) is zero, mentioned above ({\it i. e.~}
between any two successive values in the figure there is a   $q'$ that leads
to a zero expectation value). I can see here that the momentum is better
defined, even for much larger systems, but, as expected, more weight on
other $q'$ values arise for larger $N$.
I also calculated Eq.~(\ref{eq:sq}) for $N=28$ and $q=\pi/2$ but using
 $S^z_{\pi}|\psi_0\rangle$ as a target state. I find a very poorly defined
momentum centered at $q'=\pi/2$. This is expected since the reduced Hilbert
space targeted $q=\pi$ states (in addition to the ground state).

In Fig.~3, I show the dispersion
curve for 28 sites
as compared to the exact dCP dispersion (Eq.~(\ref{eq:dcp})). The difference
in the values at higher values of $q$ is due to finite-size effects.
These results are in very good agreement with those of Ref. \cite{haas}.
The inset shows the first excitation energy $\omega^1$ for $q=\pi$
as a function of $1/N$. In the thermodynamic limit this value must go
to zero. The upwards curvature for the largest systems is due to
the approximation of the method.

I find excellent agreement in excitation energies and
weights for all values of $q$ with exact results
for $N=24$ sites  \cite{takah}.

	As a check of the approximation I calculated the sum rule
\begin{equation}
\frac{1}{4\pi^2}\int_0^{\infty}d\omega \int_{q=0}^{2\pi}
S^{zz}(q,\omega)\equiv \langle \psi_0 |(S_{r=0}^z)^2|\psi_0\rangle
=\frac{1}{4}
\end{equation}
for $N=28$, 5 target states and $m=200$. I obtain a relative error of
0.86\%. I have also found that the expression for the
static structure factor $S^{zz}(q)$
given in Ref.~\cite{haldane} {\it i.e.~}
$S^{zz}(q)=-1/4 \ln(|1-q/\pi|)$, fits very accurately our data for
$N=28$ (details will be given elsewhere).

	To conclude, I have developed a method to calculate dynamical
correlation functions precisely using the DMRG technique to evaluate
the coefficients of the continuous fraction representation of the
Green's function. I show that even by considering a $0.1\%$ of the
total Hilbert space (for $N=28$ only $\sim$ 40000 states are kept)
a reasonable description of the low energy excitations is obtained.
I also show that it is possible to obtain
states with well defined momenta if the appropriate target states are used.
The numerical computation has been performed on a workstation. A better
performance (more accuratetly described excitations, larger systems)
can surely be obtained by supercomputers, where, due to a larger
memory space, a larger reduced Hilbert space can be considered. \\ \\

	I aknowledge profitable discussions with P. Horsch,
C. Balseiro and P. Fulde.

\narrowtext

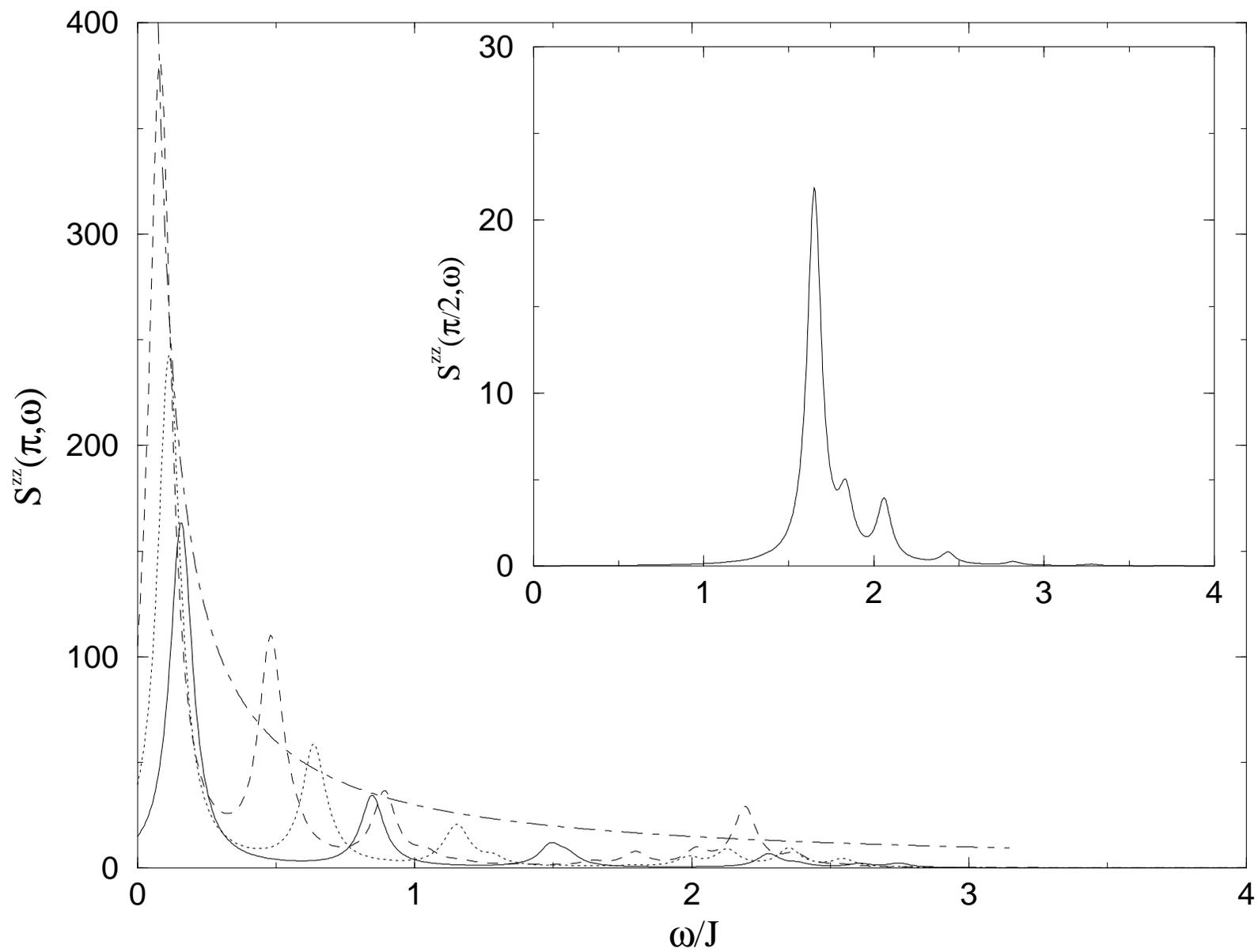
\begin{figure}
\caption{Spectral densities for $q=\pi$ and $N=28$ (continuous line),
$N=40$ (dotted line) and $N=60$ (dashed line). The analytical expression
(12) is shown with a dashed-dotted line. Inset: Spectral density
for  $q=\pi/2$ for $N=28$ ($\eta=.05$).}
\end{figure}

\begin{figure}
\caption{ Momentum weights (Eq.~(\protect\ref{eq:sq})) of a target state with
 a) $q=\pi/2$ and $N=24$ (circles),  $N=28$ (squares) and $N=36$ (diamonds);
b) $q=\pi$ for $N=28$ (circles),  $N=44$ (squares), $N=60$ (diamonds) and
 $N=72$ (triangles). The dotted lines are a guide for the eye.}
\end{figure}

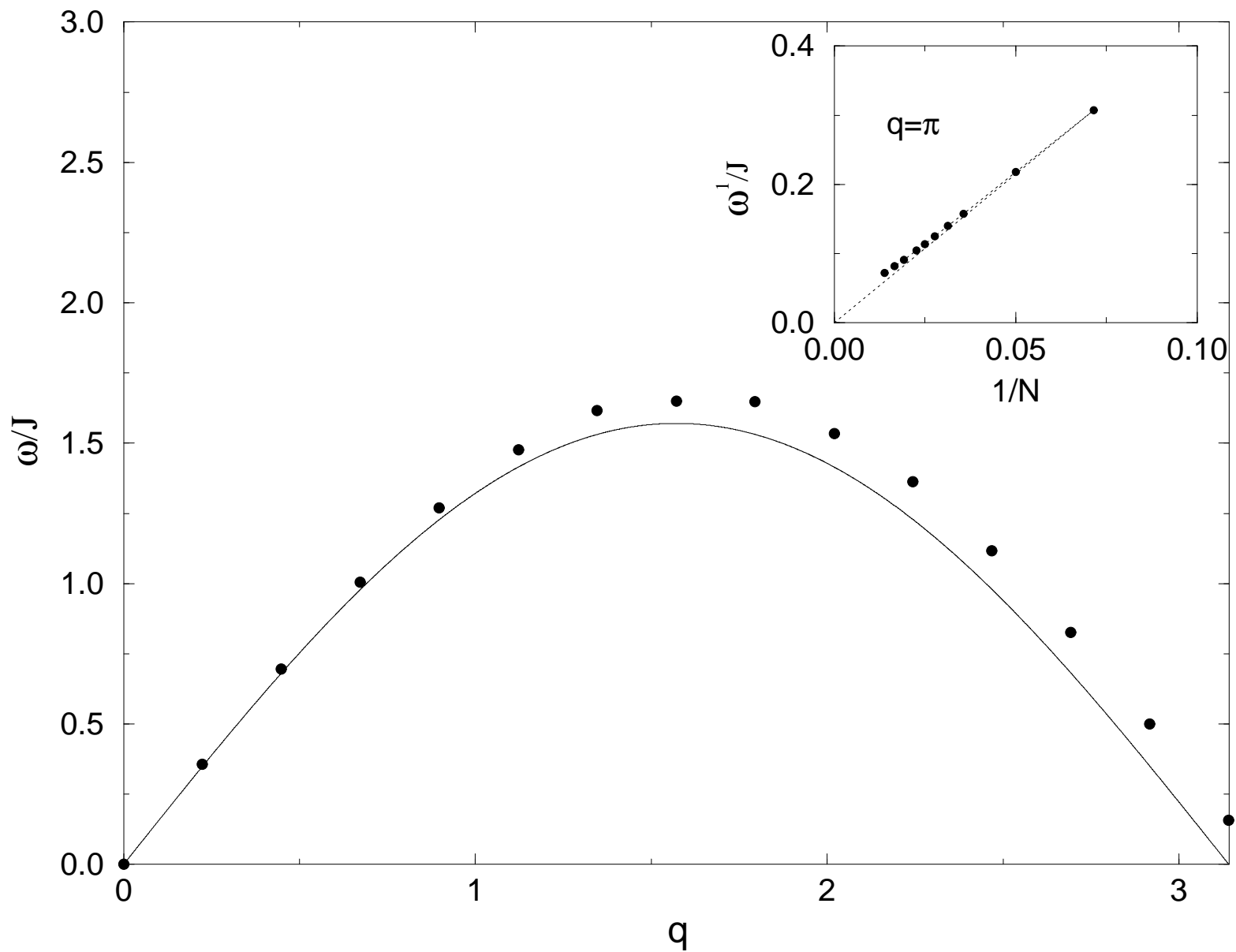
\begin{figure}
\caption{Dispersion relation of the lowest energy excitation. The full
line is the spin wave curve $\omega^l_q$ (Eq.~(\protect\ref{eq:dcp})). The dots
are our data for $N=28$ and $m=200$. Inset: First excitation energy
as a function of $1/N$ for $N=14,20,28,32,36,40,44,52,60$ and $72$. The dotted
line is a guide to the eye.}
\end{figure}

\end{multicols}
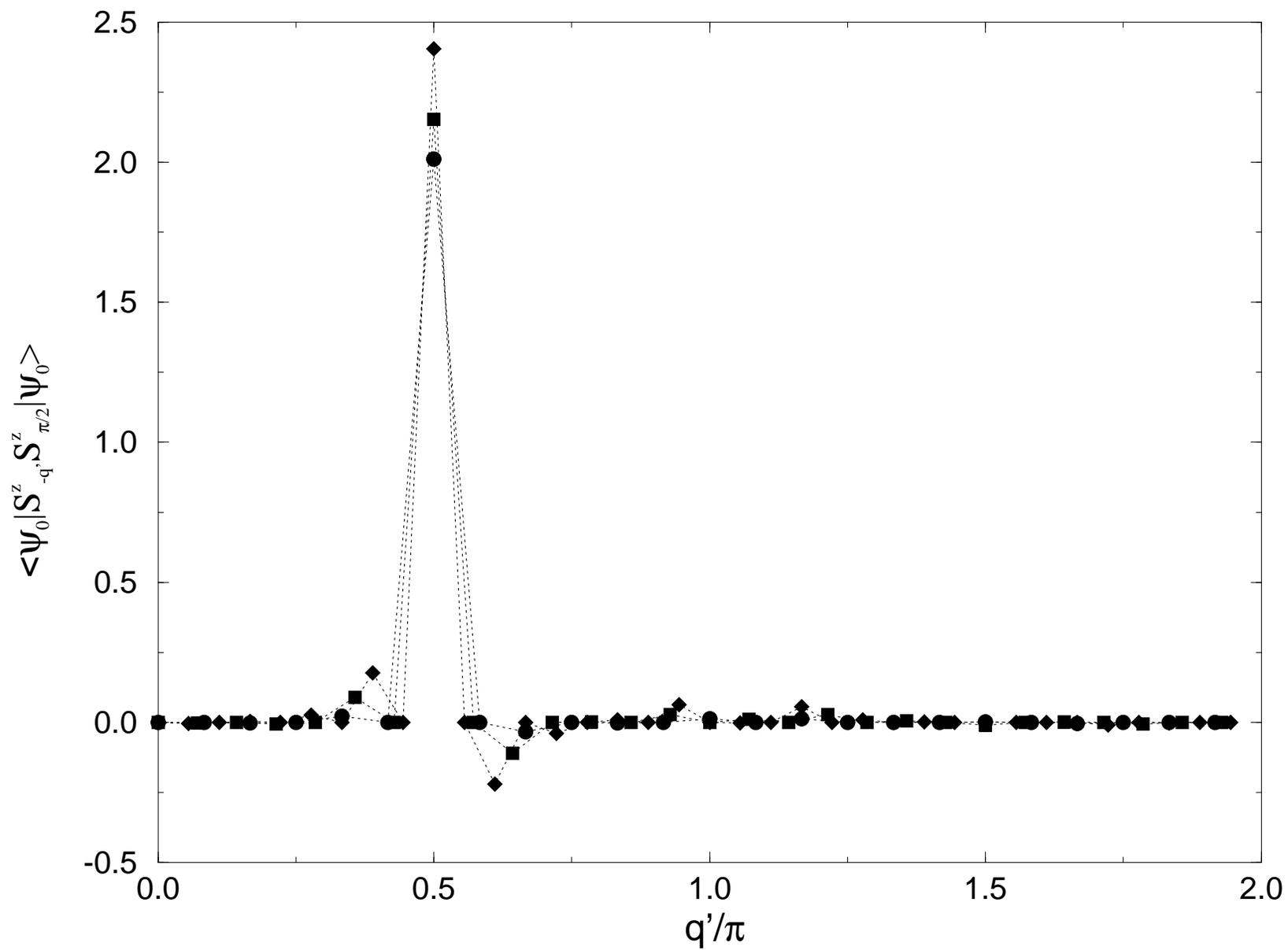
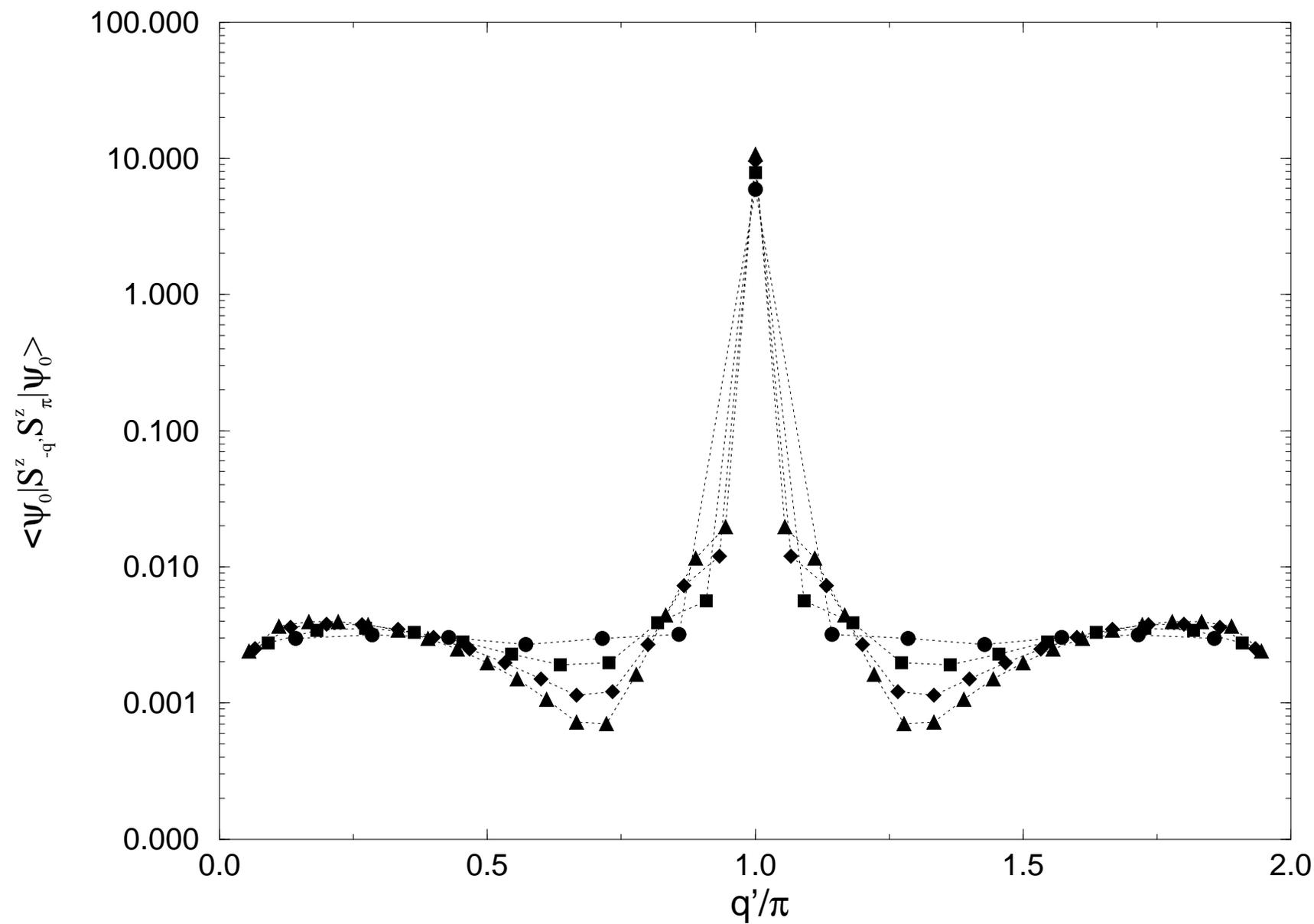
\end{document}